\documentclass[a4paper]{article}

\usepackage{xcolor}
\usepackage[T1]{fontenc}
\usepackage[french]{babel}
\definecolor{darkblue}{rgb}{0.0, 0.0, 0.55}
\definecolor{cobalt}{rgb}{0.0, 0.28, 0.67}
\definecolor{coolblack}{rgb}{0.0, 0.18, 0.39}
\PassOptionsToPackage{hyphens}{url}\usepackage[colorlinks=true, linkcolor=coolblack, urlcolor=cobalt, citecolor=coolblack]{hyperref}
\RequirePackage[hyphens]{url}
\usepackage{libertine}
\usepackage{inconsolata}
\author{Ludovic Courtès}
\date{26 juillet 2021}
\title{Reproduire les environnements logiciels : un maillon incontournable de la recherche reproductible}
\hypersetup{
 pdfauthor={Ludovic Courtès},
 pdftitle={Reproduire les environnements logiciels : un maillon incontournable de la recherche reproductible},
 pdfkeywords={},
 pdfsubject={},
 pdfcreator={Emacs 27.2 (Org mode 9.4.6)}, 
 pdflang={French}}
\begin{document}

\maketitle

\section*{Introduction}
\label{sec:orgcee7f08}

Un constat est largement partagé : puisque les logiciels font
dorénavant partie intégrante du processus scientifique, une démarche
de recherche reproductible — un pléonasme ! — doit intégrer le
logiciel.  Mais de quelle manière au juste ?

Le deuxième Plan national pour la science ouverte
\cite{mesri2021planscienceouverte}, publié en juillet 2021, « soutient »
la diffusion du code source des logiciels de recherche sous licence
libre permettant la diffusion sans restriction, mais aussi la
modification et la diffusion de versions modifiées.  C’est une
transcription naturelle du processus scientifique : le travail de
critique ne peut se faire correctement que si les pairs peuvent
étudier le code source et faire leurs propres expériences.  Le Plan
insiste aussi sur la conservation des codes sources grâce à Software
Heritage, sans quoi ce travail devient vite impossible.

Que le code source soit disponible est une condition nécessaire mais
pas suffisante.  Les sociétés savantes ont mis en place un système de
badges pour évaluer le niveau de reproductibilité des résultats
décrits dans leurs publications.  Celui de l'\emph{Association for Computer
Machinery} (ACM) dispose de trois niveaux selon que le code est
disponible (premier niveau), est utilisable (deuxième niveau), ou que les
résultats ont été reproduits indépendamment en faisant tourner le
code\footnote{\href{https://www.acm.org/publications/policies/artifact-review-badging}{Artifact Review and Badging – Version 1.0}, Association for
Computer Machinery, August 2020}.  La reproduction des environnements logiciels — le fait de
pouvoir déployer précisément l’ensemble logiciel qui a servi à une
production scientifique — est un aspect qu’on relègue volontiers au
rang de détail technique mais qui est pourtant incontournable pour
parvenir à cet objectif de reproductibilité.  Quels outils, quelles
méthodes existent pour y parvenir ?

\section*{Entre « gestion » et « gel » des environnements logiciels}
\label{sec:orge9683a6}

Beaucoup de logiciels de recherche sont développés pour GNU/Linux et
tirent parti des outils de déploiement logiciel qu’on y trouve.  Les
« distributions » GNU/Linux telles que Debian et Ubuntu se basent sur
des outils de \emph{gestion de paquets} comme \texttt{apt} qui permettent
d’installer, de mettre à jour ou de retirer les logiciels.  Ces outils
ont une vision du \emph{graphe de dépendance} des logiciels et permettent
de savoir quels logiciels sont présents sur la machine.

Malheureusement, ces outils ont deux limitations : ils requièrent les
droits d’administration système, et ils ne permettent de déployer
qu’un seul environnement logiciel à la fois.  Pour cette raison, en
particulier dans le domaine du calcul intensif (\emph{high-performance
computing} ou HPC), on a développé d’autres outils de
gestion de paquets destinés à être utilisés \emph{au dessus} celui du
système, avec l’avantage d’être utilisables sans les droits
d’administration, par chaque utilisateur·ice, qui peut ainsi déployer
ses environnements logiciels.  Les outils populaires dans cette
catégorie incluent CONDA, Spack et EasyBuild, ainsi que des outils
dédiés à un langage de programmation (pour Python, Julia, R, etc.).

L’expérience a rapidement montré que cet empilement d’outils de
déploiement devient vite préjudiciable à la reproductibilité, car
chaque outil ignore celui « du dessous », bien qu’il dépende de ce que
celui-ci a installé.  Autrement dit, chaque outil ne voit qu’une
partie du graphe de dépendance de l’ensemble.  À cela s’ajoute le fait
que, même pris individuellement, ces outils ne permettent pas ou
difficilement de reproduire un environnement logiciel à l’identique.
C’est notamment pour cette raison qu’une deuxième approche du
déploiement logiciel s’est popularisée : celle qui consiste à \emph{geler
l’environnement logiciel}, plutôt que d’essayer de le décrire.

L’idée peut se résumer ainsi : puisqu’il est difficile voire
impossible de redéployer un environnement logiciel à l’identique en
utilisant ces outils de gestion de paquets, créons l’environnement une
fois pour toutes puis sauvegardons les octets qui le composent — les
fichiers de chaque logiciel installé.  On obtient ainsi une \emph{image}
binaire, qui permet, sur n’importe quelle machine et à n’importe quel
moment, de relancer les logiciels scientifiques d’intérêt.  Le plus
souvent, on utilise les outils à base de « conteneurs » Linux, tels
que Docker ou Singularity, mais une machine virtuelle peut aussi
remplir cette fonction.

L’approche est séduisante : puisqu’on a tous les octets des logiciels,
la reproductibilité est totale ; on a la garantie de pouvoir relancer
les logiciels, et donc, de reproduire l’expérience scientifique.  Mais
l’inconvénient est de taille : puisque l’on n’a \emph{que} les octets des
logiciels, comment savoir si ces octets correspondent vraiment au code
source que l’on croit exécuter ?  Comment expérimenter avec cet
environnement logiciel, dans le cadre d’une démarche scientifique,
pour établir l’impact d’un choix de version, d’une option de
compilation ou du code d’une fonction ?  L’approche est pratique,
c’est indéniable, mais ces deux faiblesses fragilisent l’édifice
scientifique qui se bâtirait sur ces fondations.

\section*{Le déploiement logiciel vu comme une fonction pure}
\label{sec:org476c2cc}

GNU Guix\footnote{GNU Guix, \url{https://guix.gnu.org}} est un outil de déploiement logiciel qui cherche à
obtenir le meilleur des deux mondes : la reproductibilité parfaite des
environnements « gelés » dans des conteneurs, et la transparence et la
flexibilité des outils de gestion de paquets.  Il est issu de travaux
à la croisée de l’ingénierie logicielle et des langages, d’abord en
bâtissant sur le modèle de \emph{déploiement purement fonctionnel} de Nix
\cite{dolstra2004nix} et en étendant Scheme, un langage de programmation
fonctionnelle de la famille Lisp, avec des abstractions permettant
d’en tirer partie.  En termes pratiques, Guix hérite de Nix les
fondations permettant la reproductibilité d’environnements logiciels
et fournit les outils pour exploiter ces capacités sans expertise
préalable \cite{courtes2015reproducible}.

Le principe du déploiement fonctionnel est de traiter le processus de
compilation d’un logiciel comme une \emph{fonction pure}, au sens
mathématique : les entrées de la fonction sont le code source, un
compilateur et des bibliothèques, et son résultat est le logiciel
compilé.  Les mêmes entrées mènent au même résultat ; Guix s’assure
que c’est effectivement le cas en vérifiant que les compilations sont
reproductibles \emph{au bit près}, ainsi que le font d’autres projets
participant à l’effort \emph{Reproducible Builds}\footnote{\emph{Reproducible Builds}, \url{https://reproducible-builds.org}}.  Cette approche
fonctionnelle capture ainsi l’essence de la variabilité du déploiement
logiciel.  Car non : une série de noms et de numéros de version de
logiciels \emph{ne suffit pas} à rendre compte des variations qui peuvent
être introduites lors du déploiement de chaque logiciel.

Chaque déploiement logiciel est vu comme une fonction pure, la
définition est donc récursive.  Mais puisqu’elle est récursive, qu’y
a-t-il « au tout début » ? Quel compilateur compile le premier
compilateur ?  On arrive là à une question fondamentale, presque
philosophique, mais qui a un impact très concret sur la transparence
des systèmes logiciels comme l’a expliqué Ken Thompson dans son
allocution pour la remise du prix Alan Turing
\cite{thompson1984trusting} : tant que subsistent dans le graphe de
dépendance des binaires opaques dont la provenance ne peut pas être
vérifiée, il est impossible d’établir avec certitude l’authenticité
des binaires qui en découlent.

Guix s’attaque à ce problème en basant le graphe de dépendance de ses
paquets sur un ensemble de binaires pré-compilés bien identifié et le
plus petit possible — actuellement quelques dizaines de méga-octets —,
avec pour objectif de le réduire à un seul binaire suffisamment petit
pour pouvoir être analysé par un humain \cite{janneke2020bootstrap}.
C’est là un sujet d’ingénierie et de recherche à part entière.

\section*{Déclarer et reproduire un environnement logiciel}
\label{sec:orgcff4e3a}

Guix peut s’utiliser comme une distribution à part entière avec Guix
System ou alors comme un outil de déploiement par dessus une
distribution existante et donnant accès à plus de 18 000 logiciels
libres.  Il fournit une interface en ligne de commande similaire à
celle des outils de gestion de paquets : la commande \texttt{guix install
  python}, par exemple, installe l’interprète Python, la commande \texttt{guix
  pull} met à jour la liste des logiciels disponibles et \texttt{guix upgrade}
met à jour les logiciels précédemment installés.  Chaque opération
s’effectue sans les droits d’administration système.  Plusieurs
supercalculateurs en France et à l’étranger proposent Guix et le
projet Guix-HPC, qui implique plusieurs institutions dont Inria, vise
à élargir le mouvement\footnote{Projet Guix-HPC, \url{https://hpc.guix.info}}.

Voyons maintenant comment on peut concrètement, en tant que
scientifique, utiliser cet outil pour que ses expériences
calculatoires soient reproductibles.  On peut commencer par lister
dans un \emph{manifeste} les logiciels à déployer ; ce manifeste peut être
partagé avec ses pairs et stocké en gestion de version.  L’exemple
ci-dessous nous montre un manifeste pour les logiciels Python, SciPy
et NumPy :

\begin{verbatim}
(specifications->manifest
 '("python" "python-scipy" "python-numpy"))
\end{verbatim}

Il s’agit de code Scheme.  Une utilisation avancée serait par exemple
d’inclure dans le manifeste des définitions de paquets ou de variantes
de paquets ; Guix permet notamment de réécrire facilement le graphe de
dépendance d’un paquet pour le personnaliser, ce qui est une pratique
courante en HPC \cite{courtes2015reproducible}.

Supposons que l’on ait ainsi créé le fichier \texttt{manifeste.scm}, on peut
déployer les logiciels qui y sont listés — et uniquement ceux-là —
avec, par exemple, la commande suivante :

\begin{verbatim}
guix package -m manifeste.scm
\end{verbatim}

Ce manifeste, toutefois, ne contient que des noms de paquets
symboliques, et pas de numéros de version, options de compilation,
etc.  Comment dans ces conditions reproduire exactement le même
environnement logiciel ?

Pour cela, il nous faut une information supplémentaire : l’identifiant
de révision de Guix.  Puisque Guix et toutes les définitions de
paquets qu’il fournit sont stockées dans un dépôt de gestion de
version Git, l’identifiant de révision désigne de manière non ambiguë
\emph{l’ensemble du graphe de dépendance des logiciels} — aussi bien la
version de Python, que ses options des compilations, ses dépendances,
et ceci récursivement jusqu’au compilateur du compilateur.  C’est la
commande \texttt{guix describe} qui donne la révision actuellement utilisée :

\begin{verbatim}
$ guix describe
Génération 107  20 juil. 2021 13:23:35  (actuelle)
  guix 7b9c441
    URL du dépôt : https://git.savannah.gnu.org/git/guix.git
    branche : master
    commit : 7b9c4417d54009efd9140860ce07dec97120676f
\end{verbatim}

En conservant cette information adossée au manifeste, on a de quoi
reproduire \emph{au bit près} cet environnement logiciel, sur des machines
différentes, mais aussi à des moments différents.  Le plus pratique
est de stocker cette information dans un fichier, au format que Guix
utilise pour représenter les \emph{canaux} utilisés\footnote{Un \emph{canal} Guix est une collection de définitions de paquets
stockée dans un dépôt Git.} :

\begin{verbatim}
guix describe -f channels > canaux.scm
\end{verbatim}

Une personne souhaitant reproduire l’environnement logiciel pourra le
faire avec la commande suivante :

\begin{verbatim}
guix time-machine -C canaux.scm -- \
    package -m manifeste.scm
\end{verbatim}

Cette commande va d’abord obtenir et déployer la révision de Guix
spécifiée dans \texttt{canaux.scm}, à la manière d’une « machine à voyager
dans le temps ».  C’est ensuite la commande \texttt{guix package} de cette
révision là qui est lancée pour déployer les logiciels conformément à
\texttt{manifeste.scm}.  En quoi est-ce différent d’autres outils ?

D’abord, une version ultérieure de Guix peut reproduire une ancienne
version de Guix et de là, déployer les logiciels décrits dans le
manifeste.  Contrairement à des outils comme CONDA ou \texttt{apt}, Guix ne
repose pas sur la mise à disposition de binaires pré-compilés sur les
serveurs du projet ; il peut utiliser des binaires pré-compilés — et
ça rend les installations plus rapides — mais ses définitions de
paquets contiennent toutes les instructions nécessaires pour compiler
chaque logiciel, avec un résultat déterministe au bit près.

Une des deux différences majeures par rapport à l’approche qui
consiste à geler un environnement dans une image Docker ou similaire
est le \emph{suivi de provenance} : au lieu de binaires inertes, on a là
accès au graphe de dépendance complet lié au code source des
logiciels.  Chacun·e peut \emph{vérifier} que le binaire correspond bien au
code source — puisque les compilations sont déterministes — plutôt que
de faire confiance à la personne qui fournit les binaires.  C’est le
principe même de la démarche scientifique expérimentale qui est
appliquée au logiciel.

La deuxième différence est que, ayant accès à toutes les instructions
pour compiler les logiciels, Guix fournit aux usagers les moyens
d'\emph{expérimenter} avec cet ensemble logiciel : on peut, y compris
depuis la ligne de commande, faire varier certains aspects, tel que
les versions ou variantes utilisées.  L’expérimentation reste
possible.

Et si le code source de ces logiciels venait à disparaître ?  On peut
compter sur Software Heritage (SWH en abrégé), qui a pour mission rien de
moins que d’archiver tout le code source public disponible\footnote{Software Heritage, \url{https://www.softwareheritage.org}}.
Depuis quelques années, Guix est intégré à SWH de deux manières :
d’une part Guix va automatiquement chercher le code source sur SWH
lorsqu’il est devenu indisponible à l’adresse initiale
\cite{courtes2019connecting}, et d’autre part Guix alimente la liste des
codes sources archivés par SWH.  Le lien avec l’archivage de code en
amont est assuré.

\section*{Vers des articles reproductibles}
\label{sec:orgb5af54a}

On a vu le lien en amont avec l’archivage de code source, mais le lien
en aval avec la production scientifique est également crucial.  C’est
un travail en cours, mais on peut déjà citer quelques initiatives pour
construire au-dessus de Guix des outils et méthodes pour aider les
scientifiques dans la production et dans la critique scientifique.

Alors que les annexes pour la reproductibilité logicielle des
conférences scientifiques sont bien souvent informelles, écrites en
langage naturel, et laissent le soin aux lecteurs et lectrices de
reproduire tant bien que mal l’environnement logiciel, un saut
qualitatif consiste à fournir les fichiers \texttt{canaux.scm} et
\texttt{manifeste.scm} qui constituent en quelque sorte une description
exécutable de l’environnement logiciel.  Cette approche a montré ses
bénéfices notamment pour des travaux en génomique dont les résultats
demandent de nombreux traitements logiciels.

La revue en ligne ReScience C organisait en 2020 le \emph{Ten Years
Reproducibility Challenge}, un défi invitant les scientifiques a
reproduire les résultats d’articles vieux de dix ans ou plus\footnote{ReScience, \emph{Ten Years Reproducibility Challenge},
\url{https://rescience.github.io/ten-years/}}.
C’est dans ce cadre que nous avons montré comment Guix peut être
utilisé pour décrire l’ensemble d’une chaîne de traitement
scientifique, incluant le code source du logiciel dont traite
l’article, les expériences effectuées avec ce logiciel, les courbes
produites dans ce cadre, pour enfin arriver au PDF de l’article
incluant la prose et ces courbes \cite{courtes2020:storage}.  Toute
cette chaîne est décrite, automatisée, et reproductible, de bout en
bout.  Cette approche pourrait être généralisée aux domaines
scientifiques ne requérant pas de ressources de calcul spécialisées ;
nous comptons fournir des outils pour la rendre plus accessible.

La question du déploiement logiciel se retrouve également dans
d’autres contextes.  Guix-Jupyter\footnote{Guix-Jupyter, \url{https://gitlab.inria.fr/guix-hpc/guix-kernel}}, par exemple, permet d’ajouter
à des bloc-notes Jupyter des annotations décrivant l’environnement
logiciel dans lequel doit s’exécuter le bloc-notes.  L’environnement
décrit est automatiquement déployé \emph{via} Guix, ce qui garantit que les
cellules du bloc-notes s’exécutent avec les « bons » logiciels.  Dans
le domaine du calcul intensif et du traitement de données
volumineuses, le \emph{Guix Workflow Language} (GWL) permet de décrire des
chaînes de traitement pouvant s’exécuter sur des grappes de calcul
tout en bénéficiant de déploiement reproductible \emph{via} Guix\footnote{Guix Workflow Language, \url{https://workflows.guix.info}}.

\section*{Adapter les pratiques scientifiques}
\label{sec:orgd0732fd}

La place croissante prise par le logiciel dans les travaux
scientifiques, paradoxalement, avait probablement été une des causes
de la « crise » de la reproductibilité en sciences expérimentales que
beaucoup ont observée — par la perte de bonnes pratiques anciennes
telles que les cahiers de laboratoire.  Notre souhait est qu’elle
puisse maintenant, au contraire, permettre une \emph{meilleure}
reproductibilité des résultats expérimentaux, en maintenant à tout
prix la rigueur scientifique quand on arrive dans le terrain logiciel.

De même que les outils de gestion de version sont progressivement
entrés dans la boîte à outils des scientifiques comme un moyen
incontournable de suivre les évolutions d’un logiciel, les outils de
déploiement logiciel reproductible tels que Guix mériteraient faire
partie des bonnes pratiques communément admises.  Il en va de la
crédibilité de la démarche scientifique moderne.

\bibliographystyle{plain}
\bibliography{article-1024}

\section*{Copier cet article}
\label{sec:org73eb252}
Copyright © 2021 Ludovic Courtès

\begin{quote}
Cet article est mis à disposition suivant les termes de la licence
Creative Commons Attribution – Partage dans les Mêmes Conditions 4.0
International (\href{https://creativecommons.org/licenses/by-sa/4.0/deed.fr}{CC BY-SA 4.0}).  Source disponible à
\url{https://git.savannah.gnu.org/cgit/guix/maintenance.git/tree/doc/sif-2021}.

Cet article est initialement paru dans le numéro de novembre 2021 de
\emph{1024}, le bulletin de la Société Informatique de France.
\end{quote}
\end{document}